# Experimental study of mini-magnetosphere

**I F Shaikhislamov, Yu P Zakharov, V G Posukh, A V Melekhov, V M Antonov, E L Boyarintsev and A G Ponomarenko**

Dep. of Laser Plasma, Institute of Laser Physics SB RAS, pr. Lavrentyeva 13/3, Novosibirsk, 630090, Russia,

e-mail: ildars@ngs.ru

***Abstract:*** Magnetosphere at ion kinetic scales, or mini-magnetosphere, possesses unusual features as predicted by numerical simulations. However, there are practically no data on the subject from space observations and the data which are available are far too incomplete. In the present work we describe results of laboratory experiment on interaction of plasma flow with magnetic dipole with parameters such that ion inertial length is smaller than a size of observed magnetosphere. A detailed structure of non-coplanar or out-of-plane component of magnetic field has been obtained in meridian plane. Independence of this component on dipole moment reversal, as was reported in previous work, has been verified. In the tail distinct lobes and central current sheet have been observed. It was found that lobe regions adjacent to boundary layer are dominated by non-coplanar component of magnetic field. Tail-ward oriented electric current in plasma associated with that component appears to be equal to ion current in the upstream part of magnetosphere and in the tail current sheet implying that electrons are stationary in those regions while ions flow by. Obtained data strongly support the proposed model of mini-magnetosphere based on two-fluid effects as described by the Hall term.

## 1. Introduction

Mini-magnetosphere defined as a magnetosphere at ion kinetic scales is a novel subject in space plasma research. It might form around small bodies like an asteroid with a remnant magnetic field, or above magnetic anomalies like on Moon or Mars. When a stand off distance by such a field is comparable or smaller than ion inertial length or ion gyroradius interaction with plasma flow is very much different from the well-known planetary magnetospheres because of the two-fluid and kinetic effects. The studies of mini-magnetosphere were motivated by Galileo spacecraft encounter with the asteroid Gaspra in 1991 and Ida in 1993 (*Kivelson et al 1993*). However, a question of remnant magnetization of asteroids remains open. The only direct measurement was made by NEAR-Shoemaker spacecraft landing on Eros that yielded virtually zero global magnetic field (*Acuña et al 2002*). The positive observation was claimed for the asteroid Braille (*Richter et al 2001*) based on a single fly by of Deep Space spacecraft with minimum approach distance of 28 km.

Since discovery of crustal magnetization in Apollo missions, their mapping by Lunar Prospector gave ample and unambiguous examples of SW interaction with the lunar magnetic anomalies. Strong magnetic enhancements called LEMEs undoubtedly associated with crustal fields have been observed at altitudes as high as 100 km often accompanied by electron energization and wave activity (*Halekas et al 2008a*). On Moon a mini-magnetosphere might be useful as a shield against SW plasma (*Halekas et al 2008b*) and unusual albedo markings have been found around several anomalies. Since then recent missions provided new data. SELENE Explorer (*Saito et al 2010*, *Lue et al 2011*) revealed distinct magnetic reflection of SW ions over the South Pole Aitken anomaly correlated with reduction of the ions reflected by lunar surface. Besides reflected ions Chandrayaan-1 spacecraft observed above the Crisium antipode anomaly a reduction of the backscattered hydrogen atoms (*Wieser et al 2010*, *Vorburger et al 2012*) giving further evidence of a surface shielding. The Chang'E-2 spacecraft observed a reduction in proton density at altitude 100 km above Serenitatis antipode anomaly (*Wang et al 2012*).

A subject of mini-magnetosphere is of interest to future applications of magnetic field sources on-board a spacecraft. One of the schemes of crew protection from energetic galactic protons involves unconstrained dipole field with huge moment up to $10^{13} A\cdot m^2$ (*Shepherd and Kress 2007*). In another conceptual design, that of magnetoplasma sail (*Winglee et al 2000*), a much weaker dipole field is inflated by on-board plasma source. In both examples a mini-magnetosphere ~20–30 km in size is created.

Current understanding of the problem is based mostly on numerical studies by Hall MHD and hybrid codes. Models predict (*Omidi et al 2002, Blanco-Cano et al 2003*) that downstream of magnetized body (see Figure 1 below) whistler and magnetosonic wake is generated while upstream there is no ion deflection and density pile up. A shocked upstream region and a strong obstacle to SW roughly resembling magnetospheric bowshock appear only when pressure balance stand off distance is larger than the ion inertia scale (*Blanco-Cano et al 2004*). Parametric study by 3D hybrid simulation (*Fujita, 2004*) showed that the size of mini-magnetosphere is equal to MHD stand off distance when ion inertial length is small and to a closest approach distance of test particles in dipole field or Stoermer radius otherwise with sharp transition in between.

In previous paper by the authors of the present work mini-magnetosphere was studied by means of experiments, 2-D Hall MHD numerical simulation and analytical analysis (*Shaikhislamov et al 2013*). A comprehensive picture was built for the first time which explains most important features of mini-magnetosphere observed in (*Shaikhislamov et al 2013*) and in other simulations sited above. Namely, when ion inertial length is larger than pressure balance distance magnetopause shifts farther away from the dipole, jump of field at magnetopause lessens and plasma penetrates into magnetosphere to be stopped at Stoermer limit. Non-coplanar component of magnetic field is found to be behind such dramatic change. In meridian plane it is out of plane component and directed perpendicular both to dipole field and flow velocity. Experimentally observed spatial structure and independence on the sign of magnetic moment gave direct evidence that this non-coplanar field is generated by magnetopause current via the Hall term. Quantitative analytical estimates of sub-solar magnetopause position, penetration velocity and Hall field were consistent with results of numerical simulation and experimental data. Developed model explains why a mini-magnetosphere is so much different. At magnetopause the Chapman-Ferraro current generates magnetic field along its direction as described by the Hall term in the Ohm' law $\mathbf{J}\times\mathbf{B}/\mathrm{nec}$. The resulting new current system advects magnetic field, as described by the same Hall term. In steady state to cancel this additional advection plasma velocity tends to be equal to current velocity which in effect means two things. First, plasma penetrates into magnetosphere, and because the jump of kinetic pressure lessens the magnetopause position correspondingly shifts away from dipole. Second, plasma dynamics inside magnetosphere is described by a particle motion law in the dipole field. In other words, Hall currents tend to cancel electric fields so ions move only under magnetic force. In this case plasma is stopped at Stoermer particle limit. Disappearance of bowshock is explained by penetration of plasma across magnetopause. With increase of Hall currents penetration velocity also increases and, when it exceeds maximum possible velocity in magnetosheath region as determined by Rankine-Hugoniot relations, a standing shock cannot exist.

Numerical simulation in (*Shaikhislamov et al 2013*) also revealed some other counter-intuitive features of mini-magnetosphere. First, upstream Solar Wind electrons bypass magnetosphere around magnetopause boundary and don't directly penetrate inside as ions do, while inside magnetosphere a quasi-stationary population of electrons exists which neutralize ion flow. Thus, if indirect processes of exchange are slow enough, magnetospheric population of electrons might develop features distinctly different from SW and this can be of fundamental and practical interest. Second, without Interplanetary Magnetic Field the lobe magnetic field sufficiently far in the tail is dominated by out of plane Hall component rather than by the field directed tail-ward. This is of practical interest for a spacecraft crossing and data interpretation as well.

So far related observations from space have been rather scanty and incomplete. It is yet to be seen that numerical and analytical studies of the subject are tested against natural mini-magnetospheres formed by the Solar Wind around small and weakly magnetized bodies. Meanwhile, laboratory modeling proved to be useful and independent way to study the physics of mini-magnetosphere. In the present paper we continue the experimental investigation of the subject with the aim to provide new data. It is shown beyond any doubt that in the upstream and tail parts of mini-magnetosphere a global non-coplanar field exists which doesn't change sign with dipole moment inversion as components coplanar to dipole field do. This Hall field is dominant in lobe parts of the tail. The electric current associated with Hall field is close to the ion current at the upstream part of magnetosphere and, tentatively, in the tail current sheet region as well. Thus, experimental data suggest that electrons are stationary in these regions and distinct lower-hybrid oscillations observed make them particular different from other parts of magnetosphere. In general, it appears that obtained results give new and more solid and systematic evidence to the model of mini-magnetosphere proposed in (*Shaikhislamov et al 2013*).

The paper consists of four sections. In the second section experimental set up is described. The third section presents results, followed by discussion and conclusions.

## 2. Experimental set up

Throughout the paper the Geocentric Solar Magnetospheric (GSM) coordinate system is used. Experiments were conducted at KI-1 Facility (*Ponomarenko et al 2001 and references therein*). Plasma is generated by theta-pinch. Discharged hydrogen ions flow along axis of vacuum chamber 5 m in length and 1.2 m in diameter. At the chamber center magnetic dipole oriented transverse to the flow is placed. Dipole epoxy cover has the radius of 2.5 cm. Operating time of theta-pinch and dipole is $50\,\mu s$ and $200\,\mu s$ respectively. After a time of about $20\,\mu s$ following discharge a quasi steady state laboratory magnetosphere with spatial scale of the order of 10 cm in the upstream direction is formed. Geometry of experiment is shown in Figure 1 while specific conditions and parameters are listed in the Table 1. Figure 1 also shows some features that were observed in experiment and which will be discussed later.

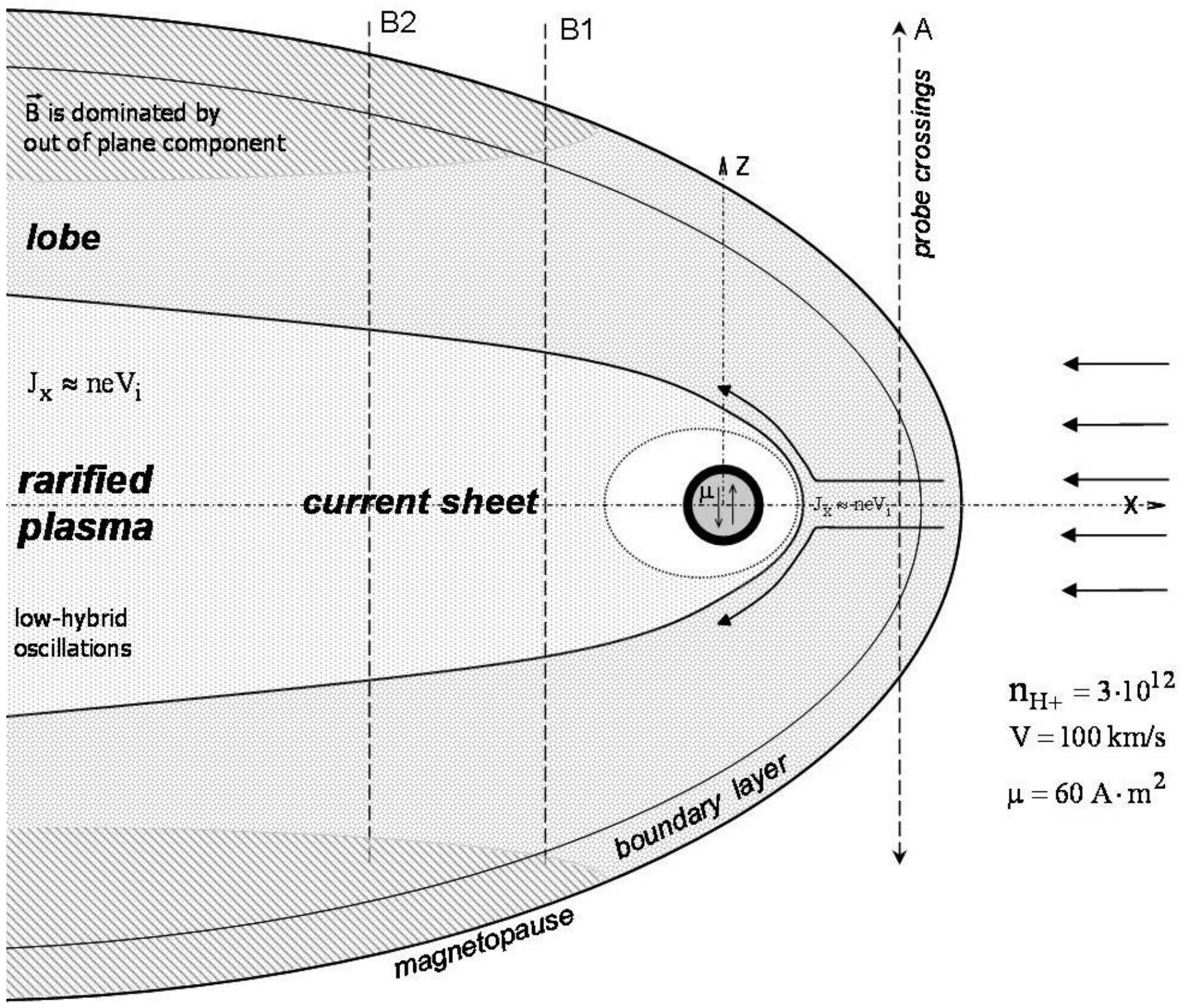

Figure 1. Experimental set up and a general structure of observed mini-magnetosphere.

Large value of plasma kinetic scales relative to the size of magnetosphere was achieved by lowering plasma density and magnetic moment. Following (*Omidi et al 2002*) we define the Hall parameter as a relation of the pressure balance stand off distance $R_M = \left( \mu^2 / 2\pi n_i m_i V_o^2 \right)^{1/6}$ to the ion inertial length $L_{pi} = c/\omega_{pi}$, $D = R_M / L_{pi}$. Note that $R_M$ is used here to define Hall parameter rather than actual size of magnetosphere. For realized experimental conditions Hall parameter was distinctly smaller than unity. Like the Solar Wind, the flow is super-sonic ($M_S \approx 3$) and sufficiently collisionless. There is no frozen-in magnetic field as analog of IMF. However, small guiding field $B_x = 5\,G$ was applied in the direction of flow, corresponding to Alven-Mach number $M_A \approx 17$ and ion Larmor radius $R_L \approx 200\,cm$. Electron kinetic scales are: electron skin $c/\omega_{pi} \approx 0.3\,cm$, Debye length $10^{-3}$ cm.

**Table 1.** Measured (left column) and calculated (right) parameters of experiment.

| parameter | value | parameter | value |
|---|---|---|---|
| density $n_{H+}$ | $3\cdot10^{12}$ cm$^{-3}$ | stand off $R_M$ | 10 cm |
| velocity $V_o$ | 100 km/s | Hall D | 0.8 |
| elec. temp. $T_e$ | 5-10 eV | Sonic Mach | ~3 |
| mag. moment $\mu$ | 60 A·m$^2$ | Knudsen Num. | >100 |
| dipole rad. $R_d$ | 2.5 cm | Reynolds Num. | ~100 |

Diagnostics consisted of three-component magnetic probes, Faraday cups and Langmuir probes described in (*Shaikhislamov et al 2012).* Spatial resolution was better than 0.5 cm. Data described in this paper have been obtained by three probe crossings of magnetosphere for negative (-Z) and positive (+Z, inverse) orientation of dipole moment. The lines of probe movement were in meridian plane along Z axis as shown in Figure 1. Due to technical reasons the range of downward motion was less than for upward.

### 3. Results

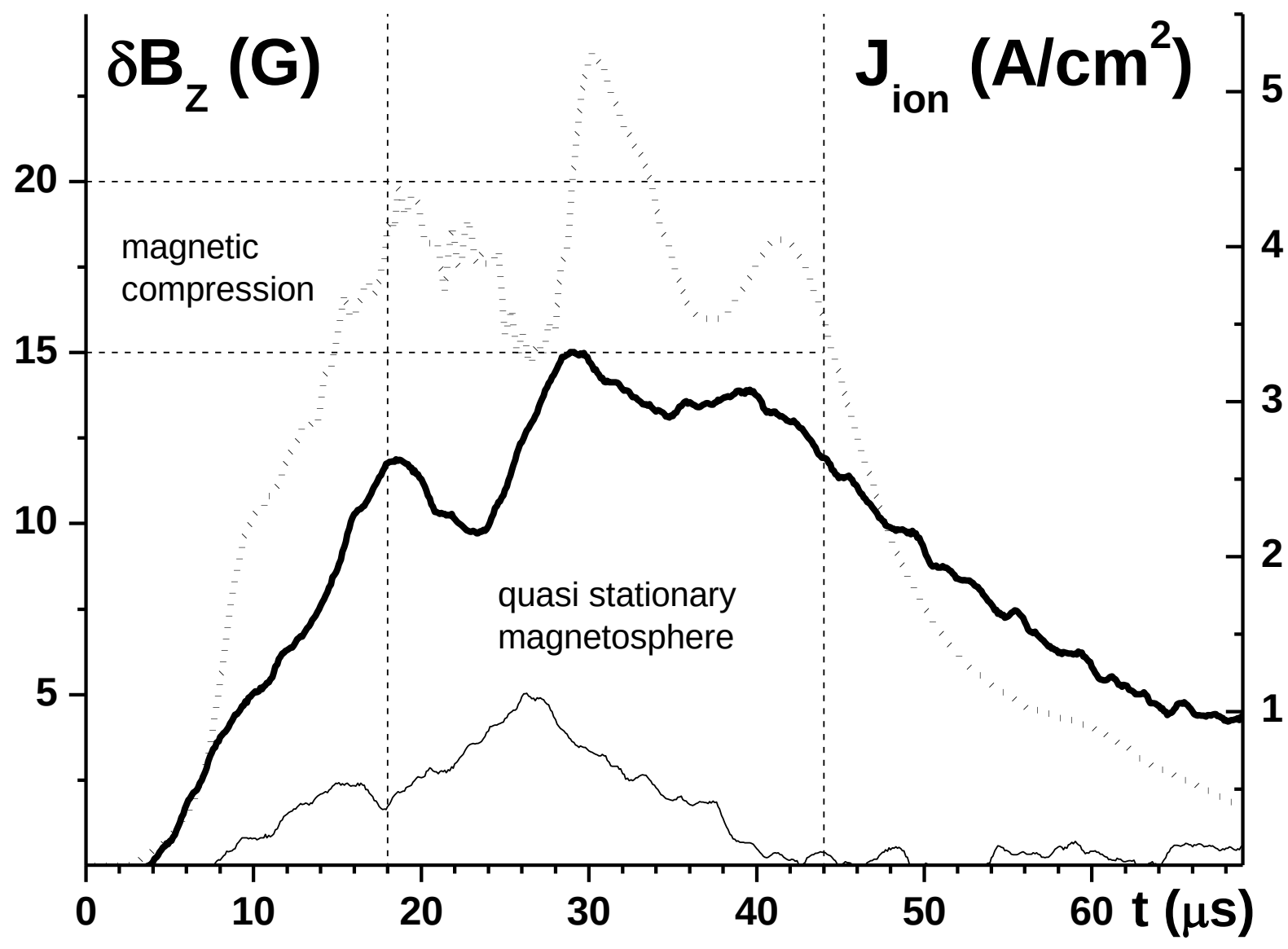

Figure 2. Time behavior of magnetic field variation (dotted line) and ion current (thin solid line) measured inside magnetosphere at position X=3 cm, Z=0. Thick solid line – ion current measured without dipole field. Straight lines indicate a time interval of quasi-stationary magnetosphere and levels of sustained magnetic compression.

Figure 2 shows typical probe signals measured close to the dipole. Magnetic probe measures magnetic variation, and total field is found by adding vacuum dipole field $B(t) = \delta B(t) + B_d$. At the X axis only Z component $\delta B_Z$ is non zero. It is produced by compression of dipole field and its dynamics corresponds to ion current $J_{ion} = neV_i$ measured when dipole has been switched off. Magnetosphere is sustained from about 20 μs to 40 μs as indicated by vertical lines which is about 20 characteristic magnetospheric times $R_M/V_o$. During this stage there are some oscillations of plasma flow and magnetic compression generated by theta-pinch discharge. When dipole is switched on ion current is strongly reduced. It means that there is a region around dipole shielded from plasma. It is indicated in Figure 1 by a white area.

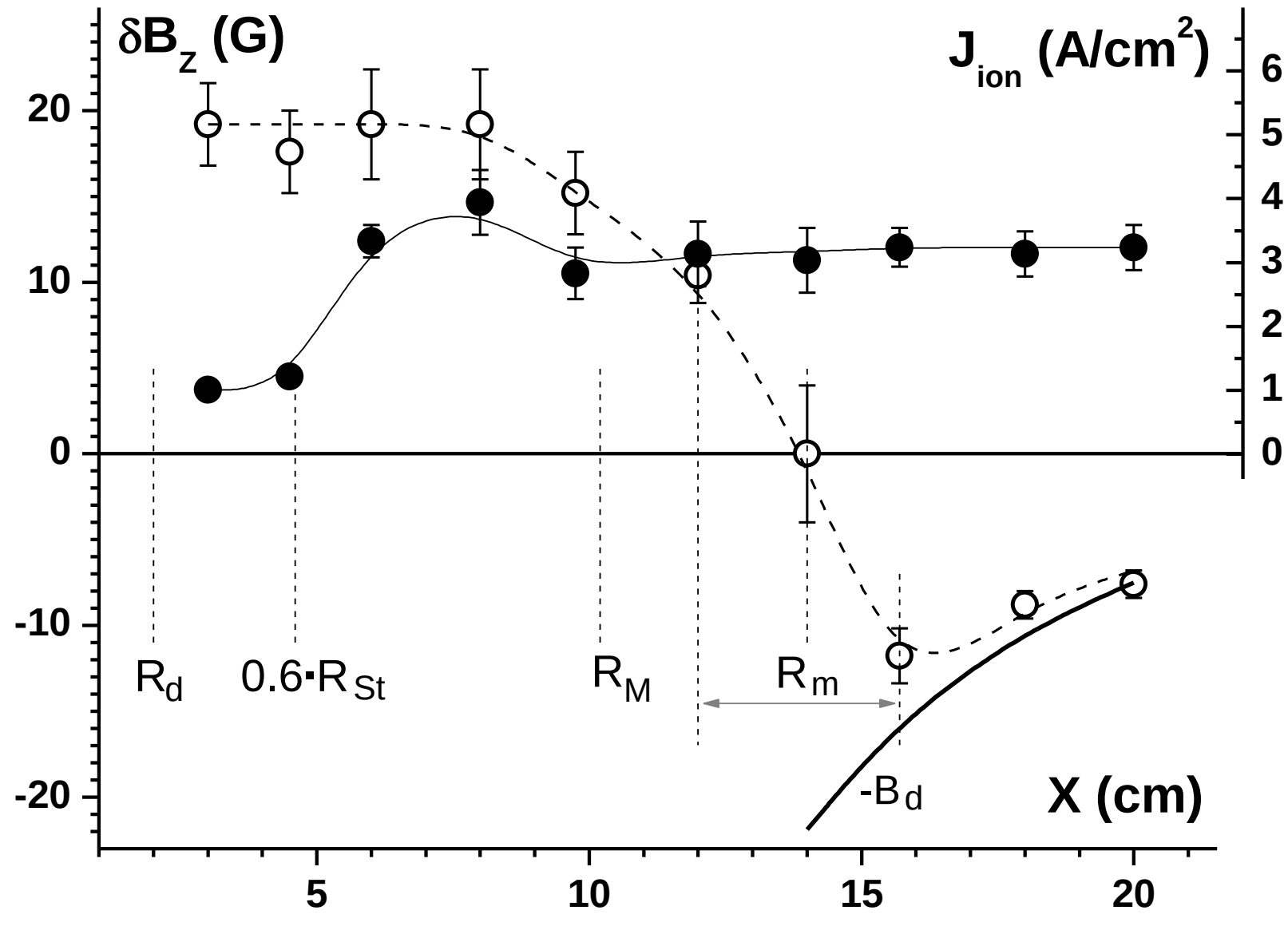

Figure 3. Profiles of magnetic field variation (O) and ion current (●) measured along X axis at Z=0 at a time of about 30 μs. Vertical lines indicate measured ($R_m$) and calculated ($R_M$) magnetopause position, boundary layer (arrow), Stoermer limit and dipole radius.

Profile of magnetosphere along X axis up to the so called "sub-solar" point and beyond is presented in the next Figure 3. Magnetic variation is approximately constant inside magnetosphere at a level of $\Delta B \approx 19\,G$, then it changes and crosses zero in the boundary layer. Away from the "sub-solar" point its value corresponds to dipole field but of opposite sign meaning that outside magnetosphere the total field $B = \delta B + B_d$ is close to zero. At a distance of about 14 cm $\delta B_Z$ crosses zero. We define such crossing as actual magnetopause position $R_m$ because it closely corresponds to maximum of magnetospheric current. At $D > 1$ thus measured $R_m$ tends to be equal to calculated stand off distance $R_M$ (*Shaikhislamov et al 2013*), while in kinetic regime, like in the described experiment, it is found much farther away. Note that the level of magnetic compression is close to the value of dipole field at the magnetopause $\mu/R_m^3 \approx 22\,G$. The jump of field across boundary layer $\approx 50\,G$ is capable to stop the flow with velocity of only 55 km/s. In other words, magnetic pressure at magnetopause is about 4 times smaller than the kinetic pressure. Thus, ions should freely penetrate through magnetopause. Probe measurements reveal that ions indeed penetrate deep inside magnetosphere. They are stopped or strongly deflected at a distance which is close to a limit of closest approach of test ions in dipole field defined in terms of a Stoermer radius $R_{St} = \sqrt{e\,\mu/V_o m_i c}$. All these features are very much similar to what have been observed by authors in previous experiments on mini-magnetosphere (*Shaikhislamov et al 2013*).

New experimental information on mini-magnetosphere comes from probe crossings along Z axis in upstream part of magnetosphere and in the tail. The upstream crossing at X=10 cm (line A in Figure 1) is presented in Figure 4. Two components are shown, $\delta B_Z$ as previously, and out-of plane component $B_Y$. There are given two sets of profiles for the normal direction of magnetic moment (negative in respect to Z axis) and inverse. One can see that profiles of $\delta B_Z$ component are bell-shaped corresponding to probe entering magnetosphere, crossing equator and exiting. $\delta B_Z$, being a compression of dipole field, changes sign with moment inversion like it should.

Out-of-plane or non-coplanar component $B_Y$ is approximately of the same value but behaves differently. First and most essential fact is that it doesn't change sign with moment inversion. Second, its spatial structure is a typical reverse profile such that positive values are above equator and negative below it. Such structure implies that there is electric current $J_X \approx -(c/4\pi)\partial B_Y/\partial z$ directed along the plasma flow. Its value at the equator is indicated in the panels and is close to the ion current $neV_i$ measured by Langmuir probe (see Figure 3). This fact and equation $J = neV_i - neV_e$ lead to the conclusion that, even if ions move inside mini-magnetosphere, the electrons are at rest. While these features have been tentatively observed in earlier paper by the authors (*Shaikhislamov et al 2013*), presented data are much more full, precise and conclusive. The time behavior of the non-coplanar component $B_Y$ is similar to that of plasma flow and is shown in Figure 5 to be compared to Figure 2.

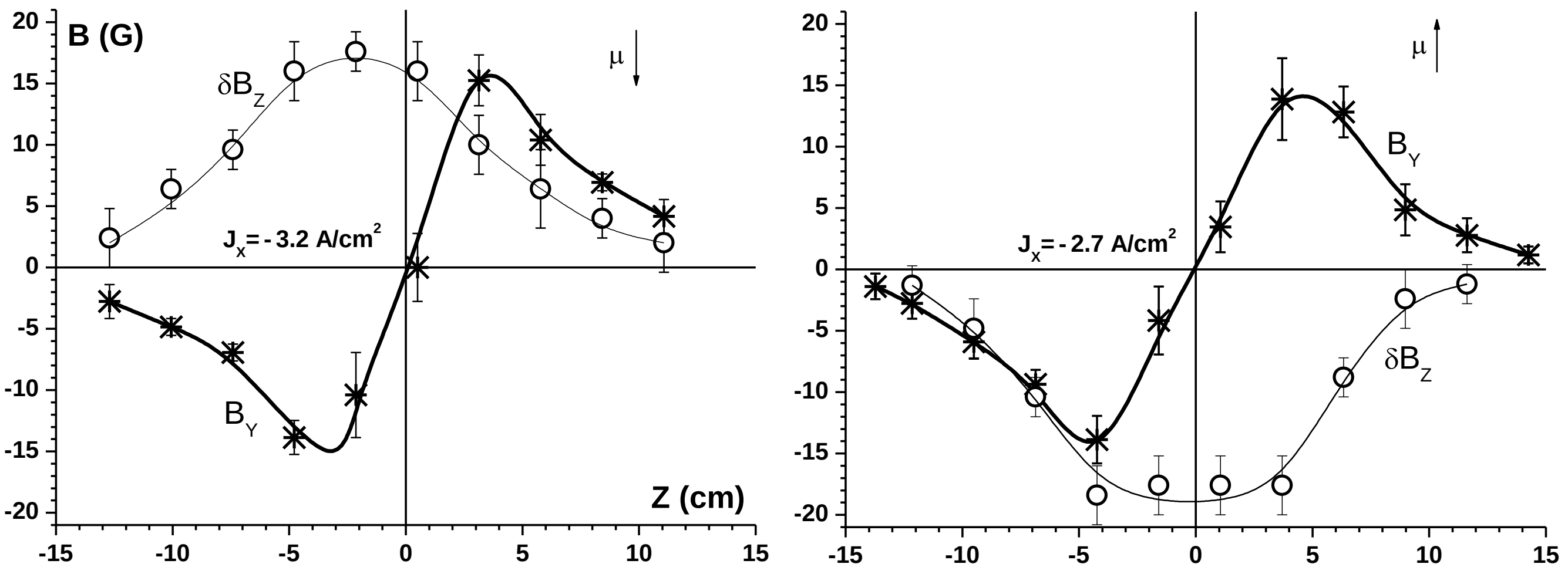

Figure 4. Profiles of $\delta B_Z$ (O) and $B_Y$ (✳) component of magnetic field measured along Z axis at X=10 cm for normal (left panel) and inverse (right panel) orientation of dipole moment.

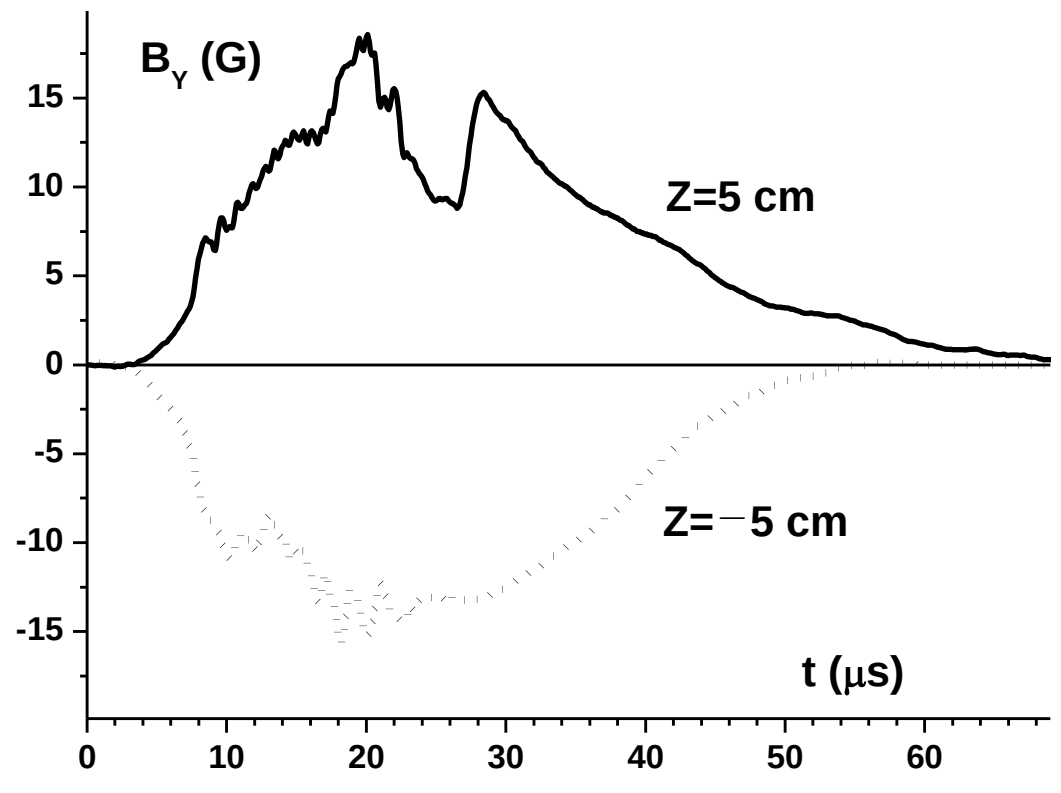

Figure 5. Time behavior of the out-of-plane magnetic component $B_Y$ above and below equator at X=10 cm.

Next crossing is in the tail at X=–10 cm (line B1 in Figure 1). In Figure 6 perturbation of all three magnetic components are shown for normal and inverse dipole moments. Once again one can see that components coplanar to dipole field $\delta B_X$, $\delta B_Z$ change sign at dipole moment inversion while non-coplanar $B_Y$ component doesn't. Near the equator plane the tail aligned component $\delta B_X$ shows sharp reversal typical of a current sheet.

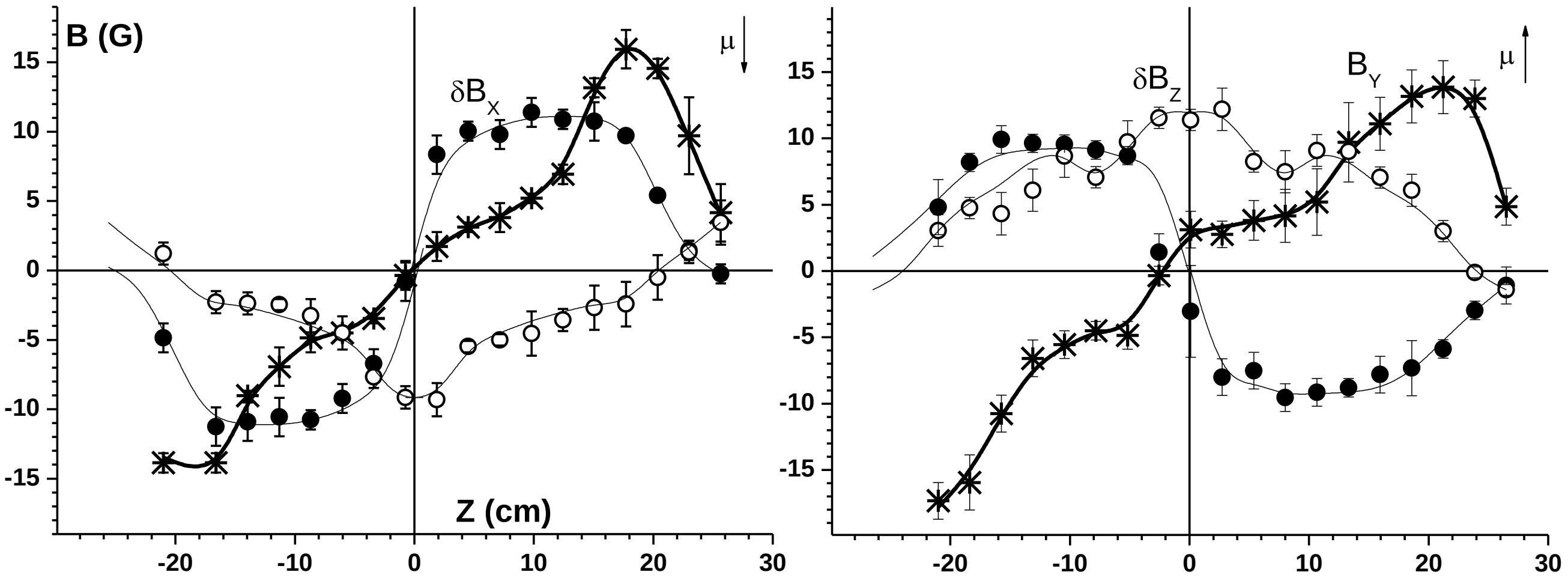

Figure 6. Profiles of $\delta B_Z$ (O), $\delta B_X$ (●) and $B_Y$ (✳) component of magnetic field perturbation measured along Z axis at X=-10 cm for normal and inverse orientation of dipole moment.

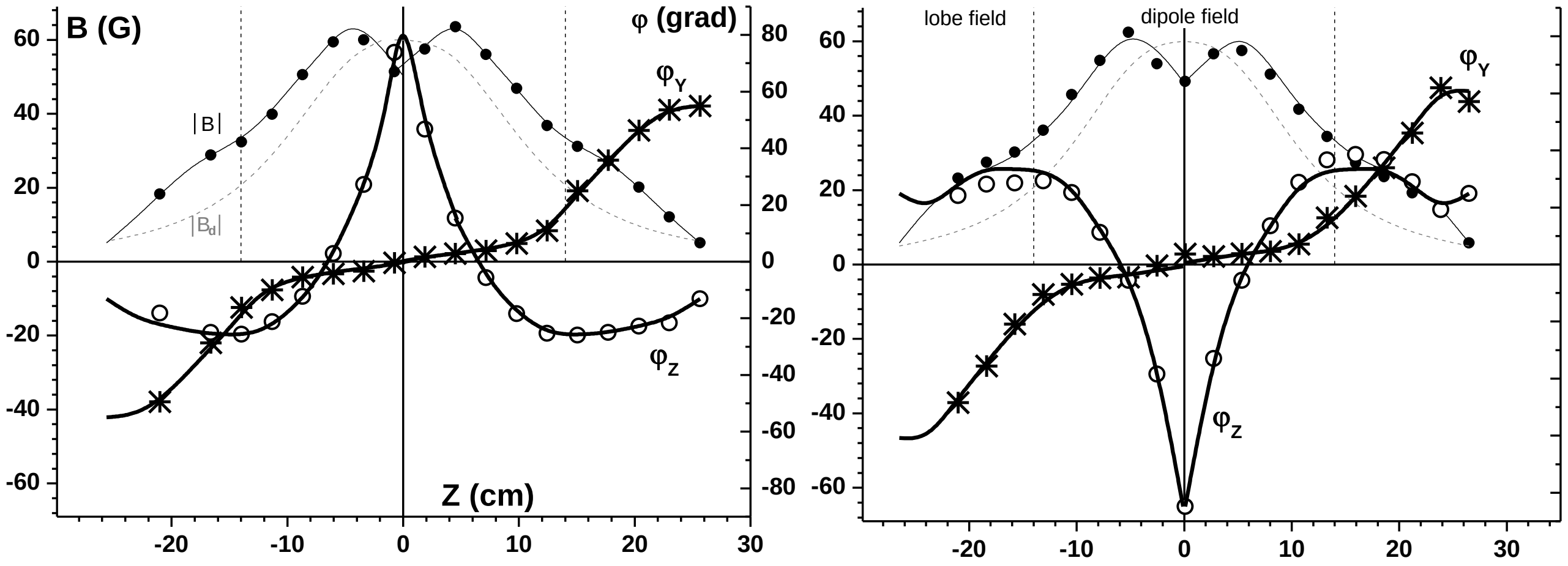

Figure 7. Profiles of total field (●) value and deviation angles $\varphi_Y$ (✳) and $\varphi_Z$ (O) of field vector measured along Z axis at X=-10 cm for normal and inverse orientation of dipole moment. Dashed line shows dipole field value.

The noticeable feature is that in the regions adjacent to magnetopause boundary out-of-plane field $B_Y$ is larger than the dipole field components. In the Figure 7 data presented in Figure 6 are plotted as total field $|B| = \sqrt{\left(\delta\vec{\mathbf{B}} + \vec{\mathbf{B}}_d\right)^2}$ and deviation of field vector from the tail orientation as two angles: $\sin(\varphi_Y) = B_Y/|B|$, $\sin(\varphi_Z) = B_Z/|B|$. At this crossing close to the dipole center the dipole field is dominant near equatorial plane. However, at about $|Z| > 14\,\mathrm{cm}$ measured total field becomes significantly larger than the dipole field. These regions can be identified as lobes. Field vector exhibits complex 3-D rotations. Immediately after crossing magnetopause it is inclined mostly along Y axis, then turns toward tail and at crossing equator performs another sharp turn in direction of dipole moment.

Prevalence of non-coplanar component in the far lobe regions becomes even more pronounced farther in the tail. In Figure 8 tail crossing at X=–20 cm is shown (line B2 in Figure 1). As dipole field at this distance is relatively unimportant, components of total field are plotted rather than components of perturbation. Note that they are the same for $B_Y$ component. Once again one can see reversal of coplanar components $B_X$, $B_Z$ and no reversal of non-coplanar $B_Y$ component with magnetic moment inversion. Note also that dipole field components dropped significantly in comparison with crossing at X=–10, while $B_Y$ decreased only slightly.

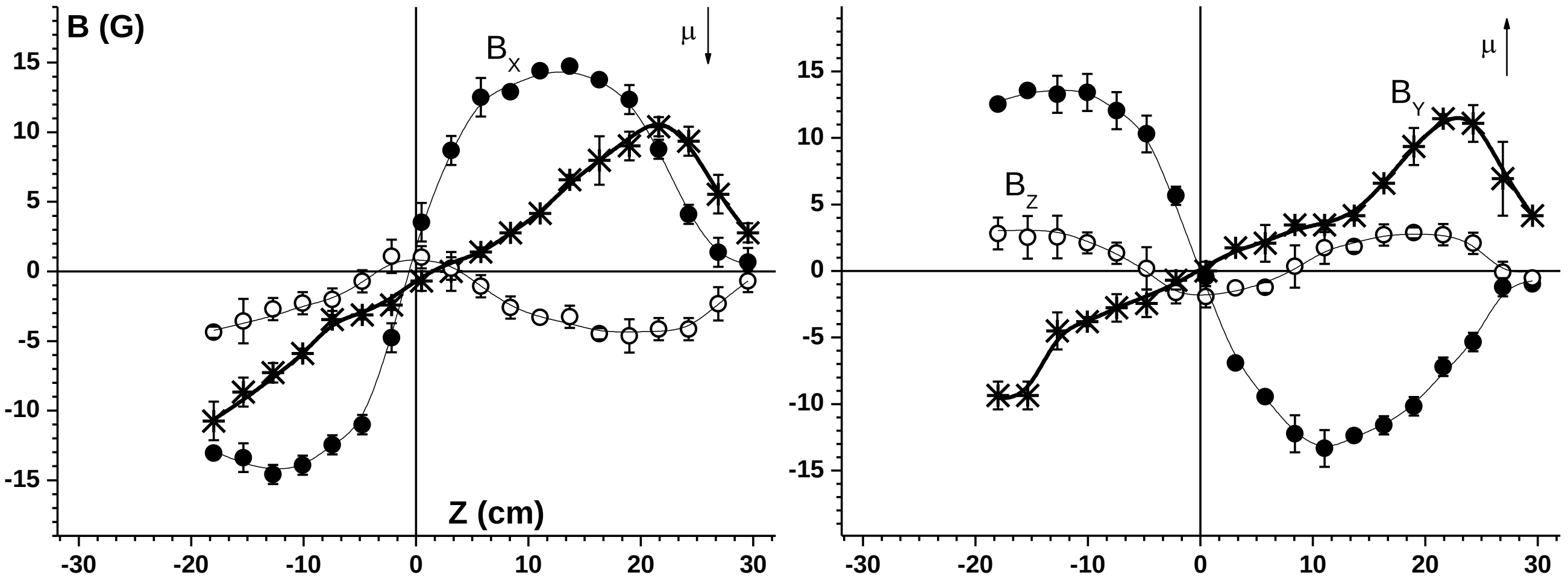

Figure 8. Profiles of $B_Z$ (O), $B_X$ (●) and $B_Y$ (✳) component of total magnetic field measured along Z axis at X=-20 cm for normal and inverse orientation of dipole moment.

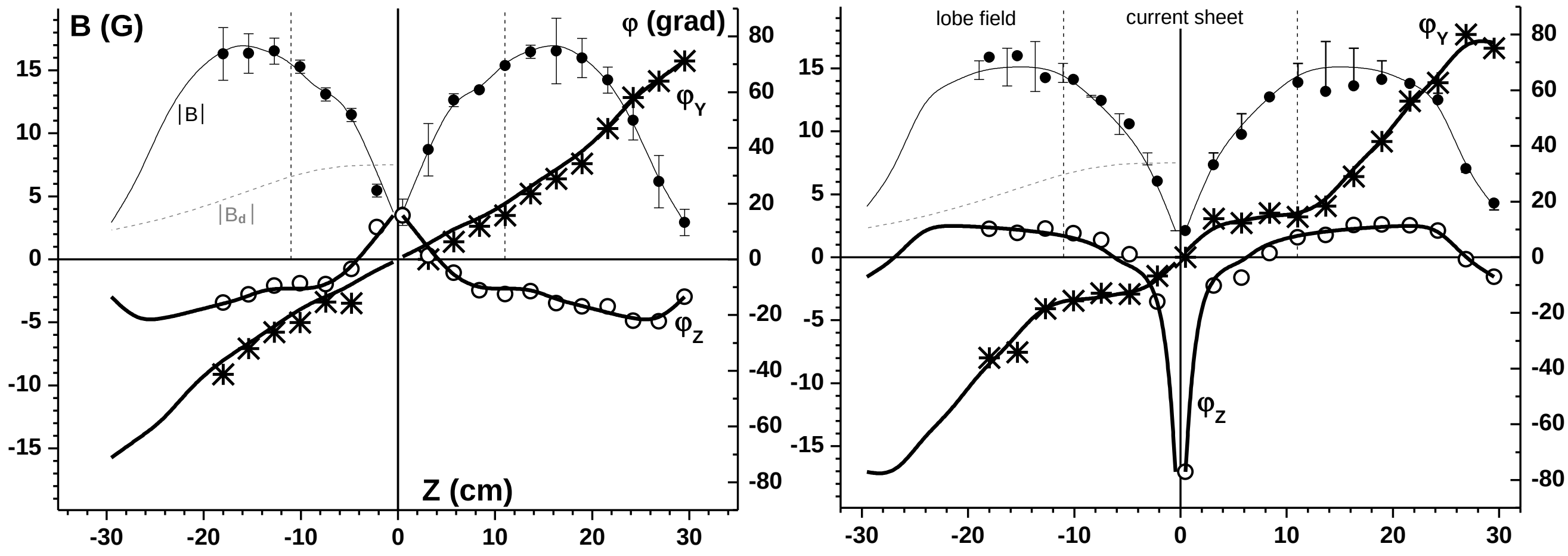

Figure 9. Profiles of total field (●) value and deviation angles $\varphi_Y$ (✳) and $\varphi_Z$ (O) of field vector measured along Z axis at X=-20 cm for normal and inverse orientation of dipole moment. Dashed line shows dipole field value.

Total field value shown in Figure 9 demonstrates that at these distances the tail structure is dominated by the current sheet. The lobe field ≈15 G is much larger than the field in the central plane ≈3 G. However, immediately outside the current sheet magnetic field vector exhibits rotation away from the tail direction and close to the magnetopause boundary it points out along direction perpendicular both to the tail and dipole magnetic moment. Evolution of the magnetic field vector can be analyzed by means of minimum variance method. For data in Figure 8 the axes of minimum variance are Z and the one deviated from the X toward Y by about 55 degrees. Result is shown in Figure 10. One can see that one of the pairs of orthogonal components reveal distinct figure 8 rotation.

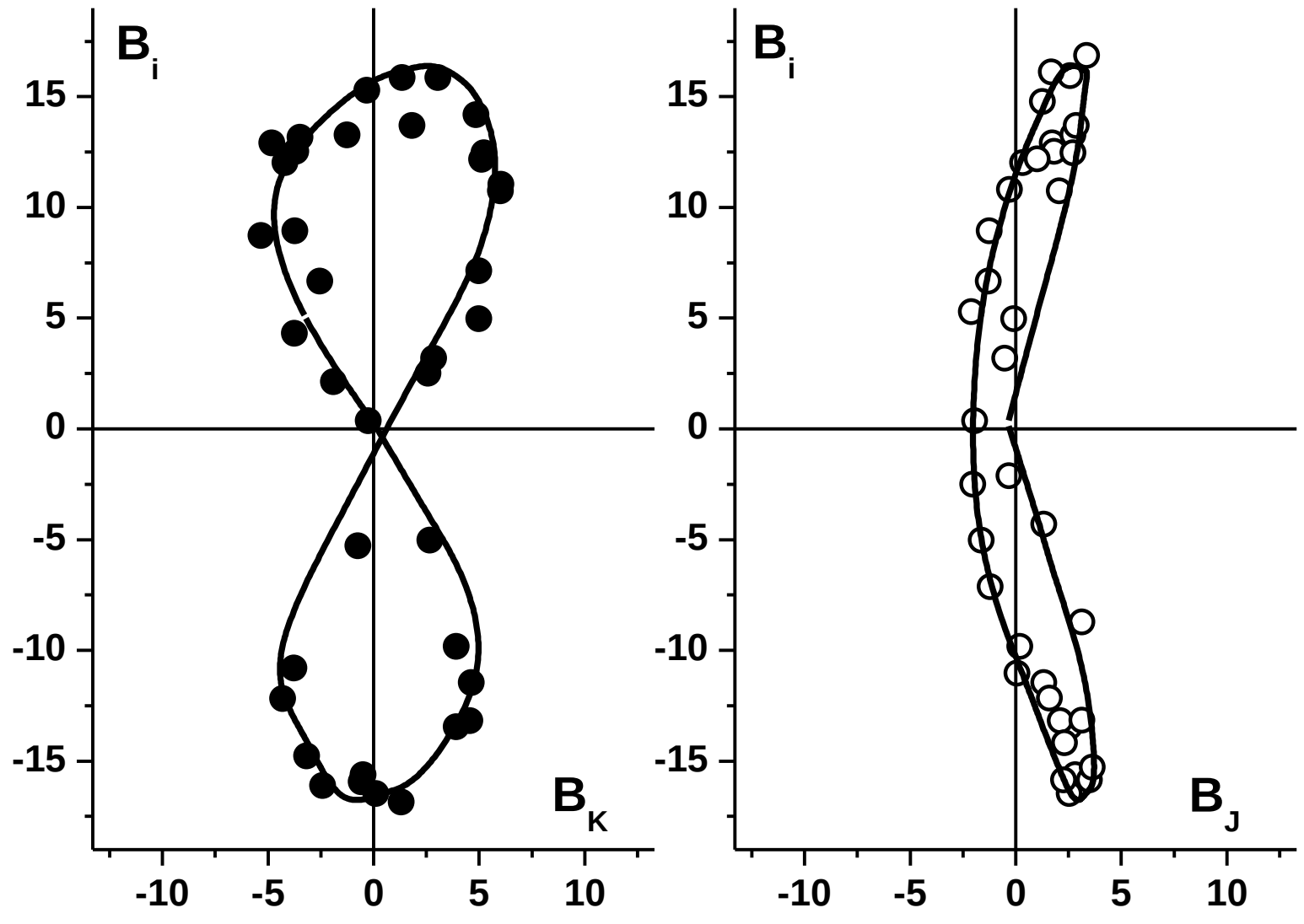

Figure 10. Minimum variance analysis for magnetic field vector plotted in Figure 8, left panel.
$B_J = B_Z, B_k = 0.57 \cdot B_X - 0.84 \cdot B_Y, B_i = 0.84 \cdot B_X + 0.57 \cdot B_Y$

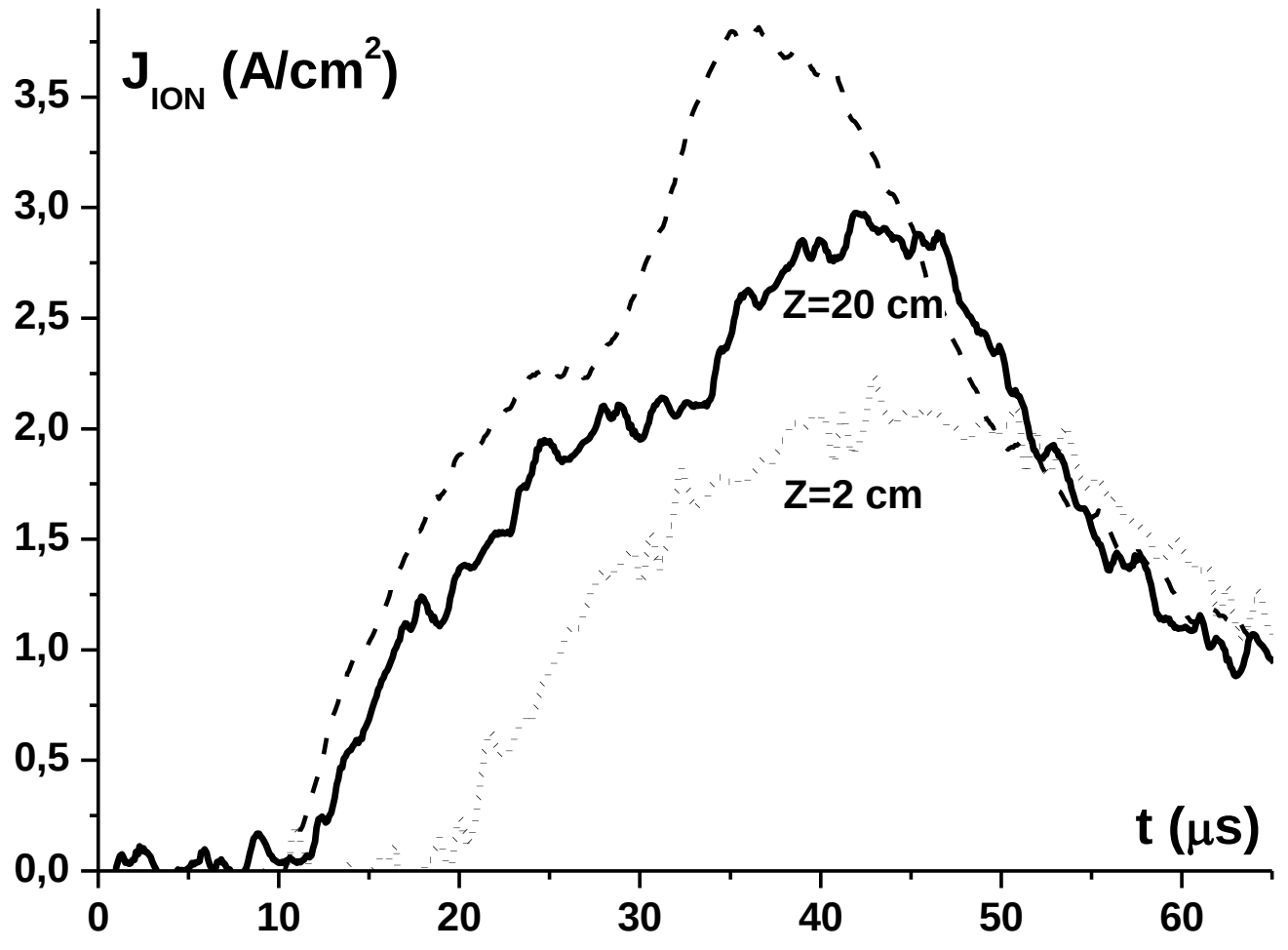

Figure 11. Dynamic signals of ion current measured upstream of dipole at X=20 cm (dashed line) and in the tail at X=-20 cm in the lobe region (solid, value amplified by 1.75) and inside the current sheet (dotted, amplified by 5).

Langmuir probe measurements reveal that the plasma is significantly rarified in the current sheet region. Its density is by order of magnitude smaller than in undisturbed space outside magnetosphere. Moreover, central sheet isn't populated by plasma directly as the lobe region is. Figure 11 demonstrates this by means of ion current signals measured in the tail at X=–20 cm, one in the central current sheet and one in the lobe. Timing of signals is compared with that measured far upstream in the undisturbed flow at X=20 cm. In the lobe region plasma arrives 3-4 µs later after passing upstream probe which more or less corresponds to upstream velocity of 100 km/s. In the central sheet however plasma arrives with large delay of ~10 µs. Note that signals in the tail are shown enlarged for easier comparison with upstream signal.

The direction of ion current can't be derived from Langmuir probe data. Nevertheless, it is reasonable to assume that this far from the dipole the plasma flows tail-ward along the X axis. In that case it can be compared with electric current component $J_X \approx -(c/4\pi)\partial B_Y/\partial z$ calculated from data presented in Figure 8. Figure 12 shows profiles of ion and electric current.

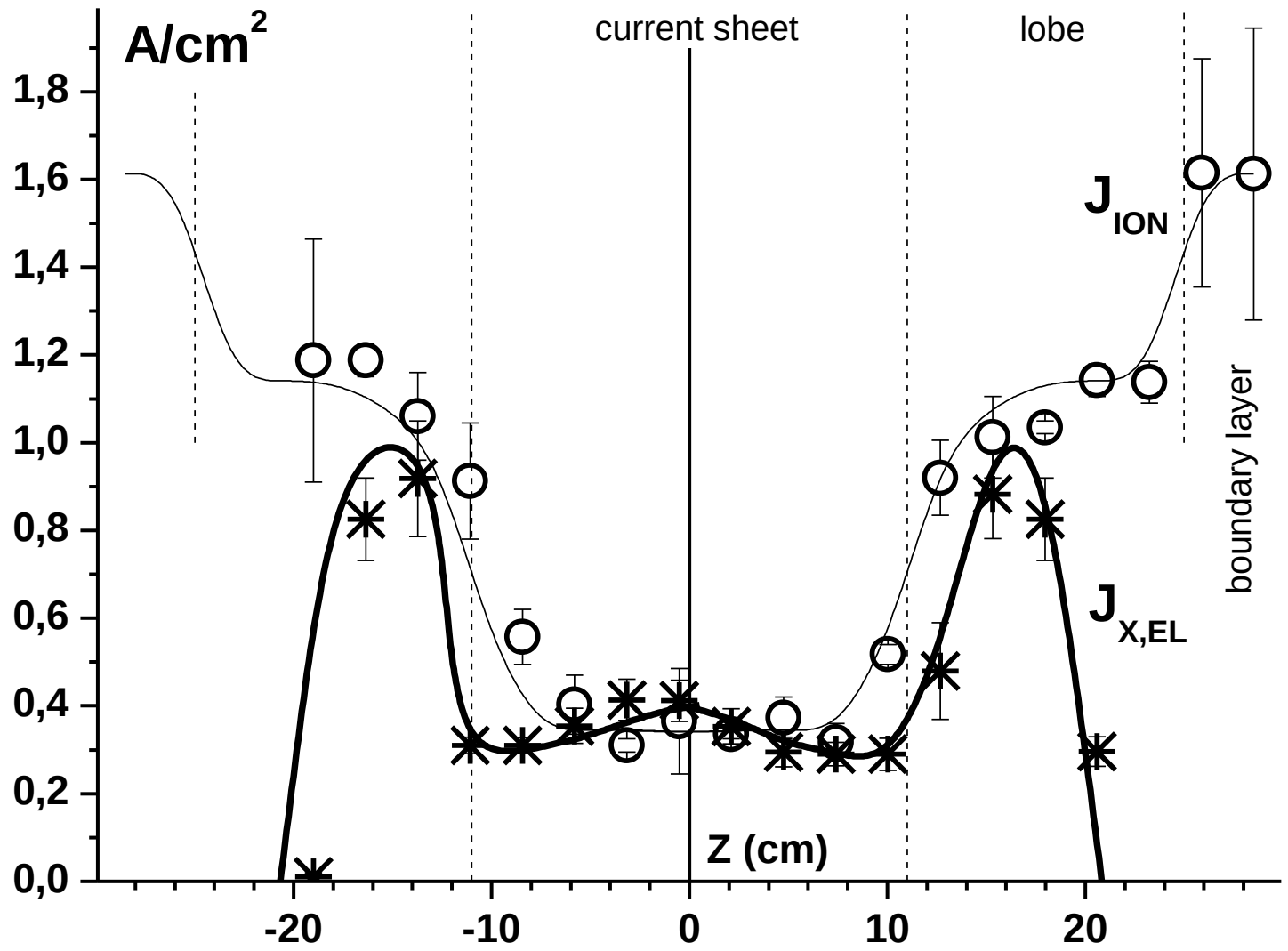

Figure 12. Profiles of ion current (O) and calculated tail-ward component of electric current (✳) measured along Z axis at X=-20 cm.

One can see that in the central sheet ion current is indeed very small. Nevertheless, its value is approximately at the same level as the X-component of electric current. They also exhibit similar sharp rise at the transition from current sheet to lobe regions. It can be tentatively concluded that in the central sheet, like at the upstream part of magnetosphere, electrons are nearly stationary. In the lobes proper electric current drops while ion current remains nearly the same. In a boundary layer near the magnetopause ion current rises again to a level of undisturbed flow which at this location was measured to be $1.65 \pm 0.25\,\mathrm{A/cm^2}$ . At crossing at X=–10 cm ion current near the central plane is also small ≈0.5 A/cm$^2$ and sharply rises at distances $|Z| \approx 10$ cm away from the equator.

Plasma in the upstream part of magnetosphere and in the tail current sheet has also another distinct feature that distinguishes it from the plasma in boundary layers and lobes. Magnetic probes register in those regions intense oscillations with frequency of order of 1 MHz which are identified as lower-hybrid oscillations. Example is shown in Figure 13. Two magnetic probe signals are given, one inside of the upstream magnetosphere and the other in the boundary layer. The first signal exhibits pronounced oscillations which appear immediately after magnetosphere formation.

Spatial distribution of oscillations measured at three crossings is presented in Figure 14. One can see that oscillations are absent or very small in boundary layers and tail lobes. In the tail they are concentrated in the current sheet region. The preferable conditions of their appearance can't be directly linked to regions with stronger magnetic field because at the X=–20 cm crossing the field in the central plane is much smaller than in the lobes while oscillations intensity behave otherwise. It can't be linked directly to plasma density as well because at the crossings at X=–10 cm and 10 cm density differs by order of magnitude while intensity of oscillations is nearly the same. The observed features are consistent with the picture that in the upstream magnetosphere and the rarified region of tail a quasi-stationary electron population forms which is not directly linked with upstream electrons like the boundary and the lobe regions do.

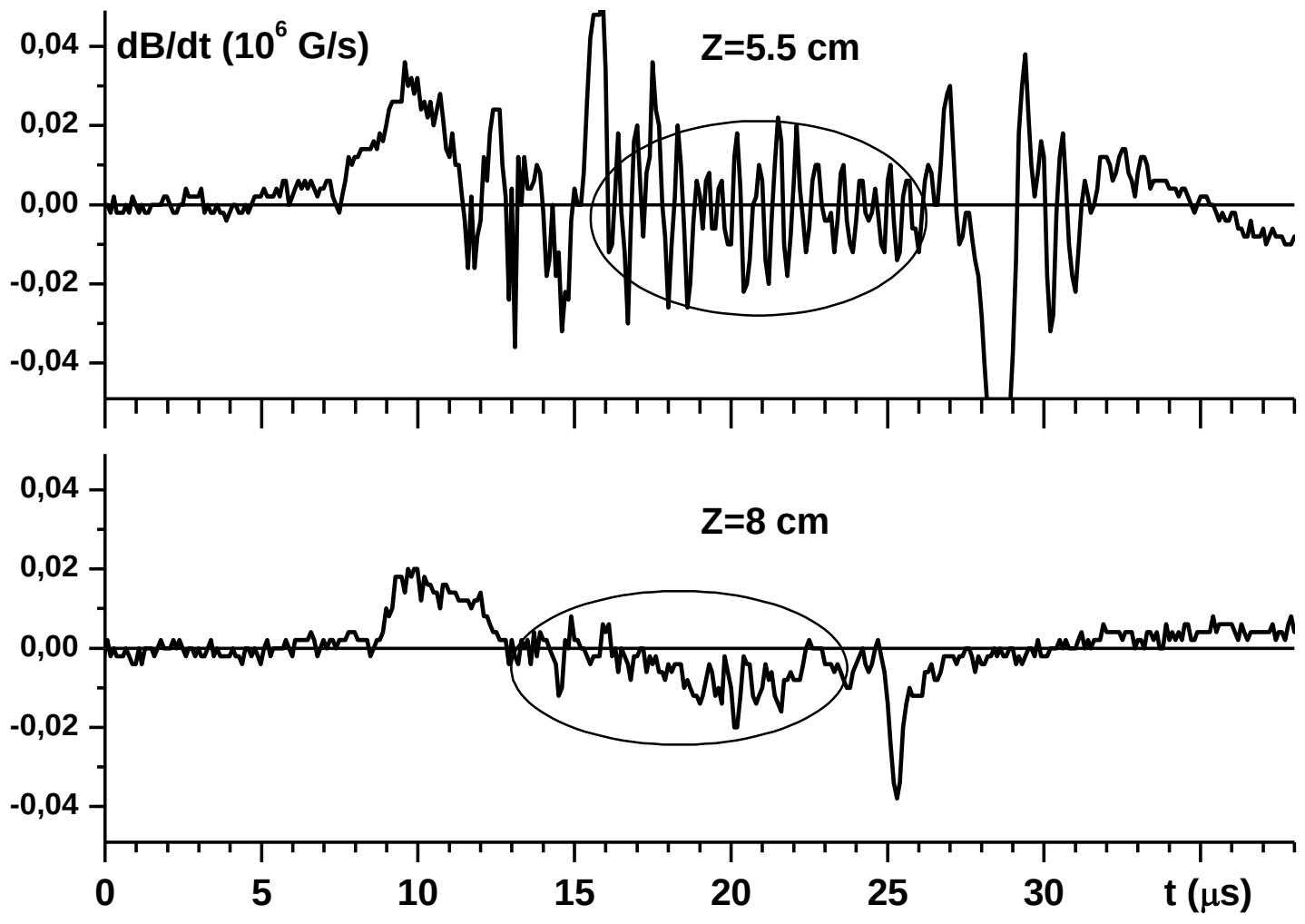

Figure 13. Magnetic signals registered at two locations inside magnetosphere both measured at X=10 cm. Circles mark periods when oscillations are observed.

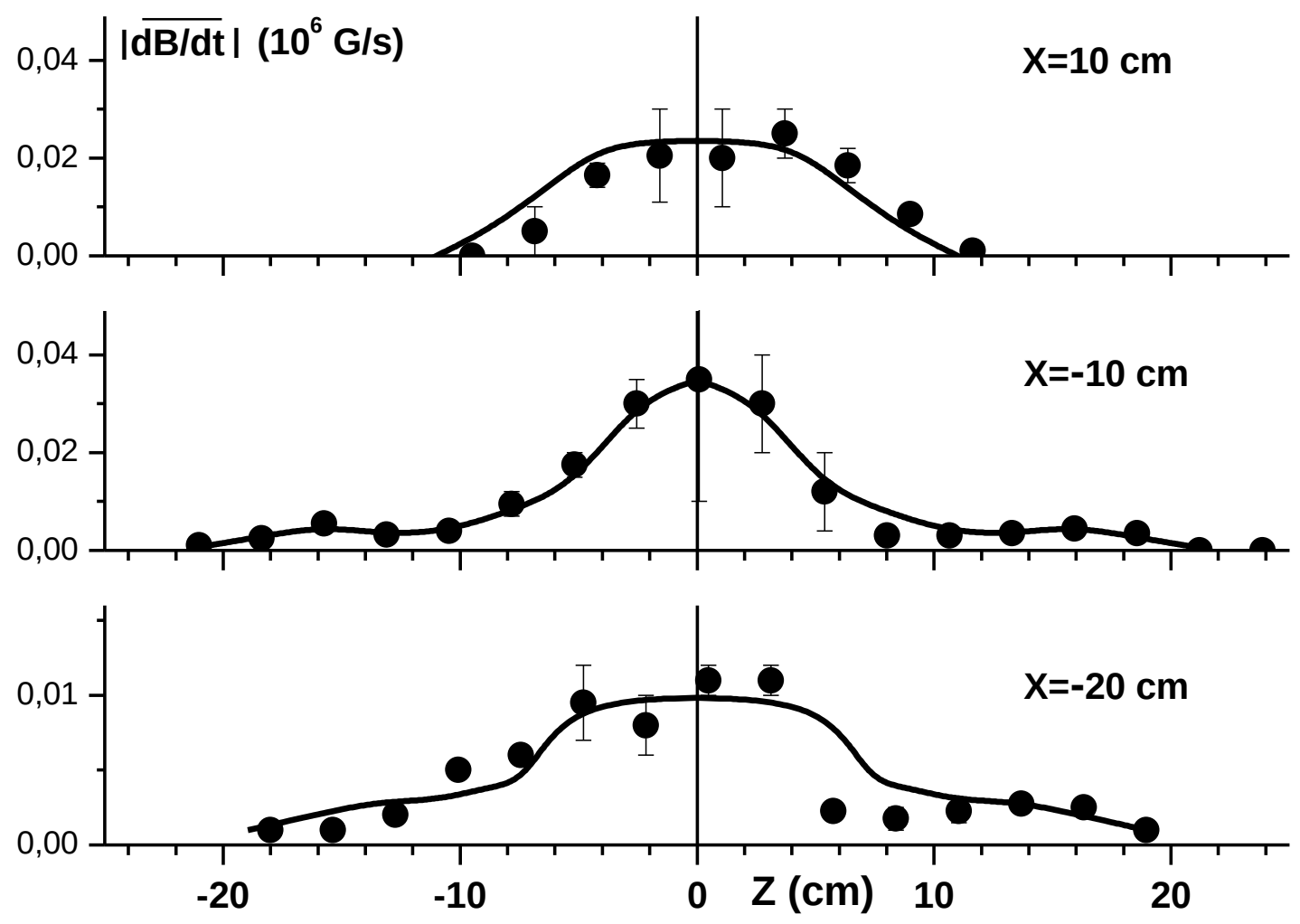

Figure 14. Spatial distribution of oscillations obtained at three crossings of magnetosphere. Values are calculated as an averaged module of oscillating part of magnetic signal.

## 4. Discussion and conclusions

In the paper new experimental data on mini-magnetosphere have been presented which can be summarized as following. In the upstream part typical features of mini-magnetosphere have been identified – the magnetopause position quite farther than the calculated stand off distance, the jump of field at magnetopause significantly less than necessary to stop the plasma flow, penetration of plasma deep inside the magnetosphere up to the Stoermer particle limit. Crossing the inner part of magnetosphere along Z-axis revealed the existence of global non-coplanar $B_Y$ component of magnetic field. Systematic measurements give unequivocal evidence that it is positive above equator and negative below it independently of the orientation of dipole magnetic moment. Moreover, the electric current associated with it is close in value to ion current and has the same direction. These prove that the Hall term is responsible for generation of $B_Y$ component and that penetration of ions deep inside magnetosphere without corresponding advection of dipole field is explained by the fact that electrons are stationary there.

Crossings behind the dipole revealed that mini-magnetosphere possesses distinct tail structure consisting of lobes where magnetic field largely exceeds dipole field and of current sheet with rarified plasma. The unusual feature is that the non-coplanar component becomes a main field component in the lobe regions adjacent to the magnetopause. Field vector performs a specific rotation around the tail axis which can be used for data interpretation. At relatively close fly-by in the tail of mini-magnetosphere one should see either a usual reversible lobe field oriented along the tail, or almost $90^o$ rotation of the lobe field before crossing the central current sheet. This depends on weather the fly-by is in equatorial or meridional plane.

It appears that in the central current sheet the tail-ward component of electric current is close to ion current like in the upstream part of mini-magnetosphere. Thus, electrons may be stationary in this region as well. This conclusion is also supported by observation that central region of the tail is populated by plasma indirectly. The other similarity between tail current sheet and the upstream part of magnetosphere is a presence of intense lower-hybrid oscillation.

Comparing presented experimental data with results of numerical 2D Hall MHD simulation of mini-magnetosphere without IMF and analytical model of (*Shaikhislamov et al 2013*) one finds very good qualitative agreement. Model and simulation predicts that upstream part of magnetosphere and central current sheet regions are shielded from SW electrons and contain quasi-stationary electron population while ions freely penetrate into those regions. SW electrons flow around the magnetosphere along magnetopause and deeper in the tail penetrate into boundary layer and lobe regions. This explains why lower-hybrid oscillations which are excited by relative motion of electrons and ions across magnetic field have been observed only in the upstream part and central sheet. In the boundary layers and lobes electrons tend to move with ions and are rapidly advected tail-ward from the regions where instability may operate. Simulation also predicts that without IMF the non-coplanar component of magnetic field is dominant in the tail lobes. This is unequivocally verified in experiment as well.

At the moment obtained data may have direct application only in future if asteroids with remnant magnetization are found or if concept of magnetic shielding of spacecraft is put in practice. In regard of magnetic anomalies on Moon or Mars situation is more complex than conducted experiment. Crustal fields are spatially irregular and have multi-pole structure. The problem is complicated as well by presence of large fraction of reflected ions, the origin and role of which is not well understood and their pick up by SW (*Kallio et al 2012*). However, the new physics found in studies of mini-magnetospheres around magnetized asteroids will help to understand the processes taking place above magnetic anomalies. Finally, experiments described in (*Shaikhislamov et al 2013*) and in the present paper open up new avenues of research. It is interesting to find out the structure of mini-magnetosphere in a terminator plane because it should have strong asymmetry in non-coplanar direction due to ion gyrorotation. Of practical interest would be modeling some aspects of Lunar magnetic anomalies, such as combination of dipoles placed below dielectric surface.

**Acknowledgements**

This work was supported by SB RAS Research Program grant II.8.1.4, Russian Fund for Basic Research grant 12-02-00367, OFN RAS Research Program 15 and Presidium RAS Research Program 22.